\documentclass[aps,
prd,
preprint,
eqsecnum,
amsmath,
amssymb
]{revtex4}

\usepackage{hyperref}
\usepackage{color}
\usepackage{xcolor}

\begin{document}

\title{Constrained field theories on spherically symmetric spacetimes with horizons} 

\author{Karan Fernandes} \email[email: ]{karan12t@bose.res.in}
\affiliation{S N Bose National Centre for Basic Sciences, Block JD, Sector III, Salt Lake, Kolkata 700106, India}

\author{Suman Ghosh} \email[email: ]{smnphy@gmail.com}
\affiliation{Tata Institute of Fundamental Research, Homi Bhabha Road, Mumbai 400005, India}

\author{Amitabha Lahiri} \email[email: ]{amitabha@boson.bose.res.in}
\affiliation{S N Bose National Centre for Basic Sciences, Block JD, Sector III, Salt Lake, Kolkata 700106, India}

\begin{abstract}
We apply the Dirac-Bergmann algorithm for the analysis of 
constraints to gauge theories defined on spherically symmetric black hole  
backgrounds.  We find that the constraints for a given theory 
are modified on such spacetimes through the presence of 
additional contributions from the horizon. As a concrete example, we 
consider the Maxwell field on a black hole background, and determine 
the role of the horizon
contributions on the dynamics of the theory. 
\end{abstract}


\maketitle

\section{Introduction}
Black holes are perhaps the most curious objects described by physics. Their construction requires
only our concepts of space and time~\cite{Chandrasekhar:1985kt}, and they are completely described
by only a few parameters, such as their mass, charge and spin. In this, black holes are exactly like 
elementary particles. Another property that black holes share with fundamental particles is our
complete lack of knowledge about their internal structure, including whether any such structure
exists. But the source of this ignorance appears to be different for the two kinds of objects.
Elementary particles are point objects which cannot be probed further, since that would require
infinite energy of the probe. A black hole on the other hand presents to the universe a closed surface
of finite size, but it is impossible to observe anything about its internal structure, as no information
passes from the inside of this surface to the outside, at least classically. 

The startling discovery by Hawking that stationary black holes radiate like a black body 
with a finite surface temperature~\cite{Hawking:1974sw}, following Bekenstein's suggestion 
that standard laws of thermodynamics applied to a black hole provided we assume that its 
entropy is proportional to the surface area of its horizon~\cite{Bekenstein:1973ur}, implies 
the possibility that a black hole has associated with it a very large number of microscopic 
states. It is natural to think that these states are in some way related to the degrees of 
freedom of the horizon. In the membrane paradigm, one replaces the black hole by a 
membrane with certain classical properties at the stretched horizon, i.e. a small distance 
outside the event horizon (an excellent overview is provided by the collection of articles 
in~\cite{Thorne:1986iy}). This is a sensible description from the perspective of an external 
stationary observer, who finds that particles cannot classically leave the interior of the 
black hole or reach the horizon from the outside in finite time. 

Thus it seems that the classical or semi-classical dynamics of fields, or gravity, on a 
spacetime which includes a horizon, may be studied by looking at the bulk and the horizon, 
and completely ignoring what happens in the interior of horizon. It has been suggested that 
in this view it should be sufficient to consider fields on a manifold with boundary. For 
gravity, this approach leads to a quantum description in which an infinite set of 
observables are localized on the boundary~\cite{Balachandran:1994up, Balachandran:1995qa, Carlip:1998wz, Carlip:1999cy}. Recently there has been a resurgence of interest in 
studying the behaviour of quantum fields near black hole horizons, motivated by various 
paradoxes and puzzles~\cite{Almheiri:2012rt, Braunstein:2009my}. The near horizon behaviour 
of fields have also been investigated in the space time of isolated 
horizons~\cite{Ashtekar:1998sp, Ashtekar:1999yj, Ashtekar:2000sz, Ashtekar:2001is, 
Booth:2000iq, Booth:2001gx}. 

Boundary conditions on the classical fields play a crucial role in all these investigations. 
In most of these papers, though not in all of them, Dirichlet (or Neumann) conditions are imposed 
on the boundary, i.e. fields (or their derivatives) are set to vanish on the (stretched) horizon. 
This is a
convenient choice for most calculations, but somewhat of an overkill, since the (stretched) 
horizon is not a physical boundary of spacetime. In particular, fields need not vanish on 
the horizon -- only invariants constructed out of the stress energy tensor need to remain 
finite. For that it is sufficient for invariants made out of the physical fields to remain 
finite on the horizon. Gauge theories are even more special in this regard, since components 
of gauge fields are not physical, but defined up to gauge transformations. So strictly speaking
it is not necessary to impose finiteness on components of gauge fields on the horizon. 

Gauge theories are characterized by the presence of redundant degrees of 
freedom, which leads to the presence of constraints, usually relations among
the corresponding momenta. The formalism for studying
the dynamics of constrained systems was discovered by Dirac~\cite{Dirac:1950pj}
and independently by Bergmann et al.~\cite{Bergmann:1949zz,Anderson:1951ta}, and has 
been applied 
to numerous theories of interest over the years~\cite{Dirac-lect-1964,Henneaux:1992ig}. 
In this paper, we will be concerned with the 
classical dynamics of gauge theories defined on spherically symmetric 
curved backgrounds, with horizons for boundaries. 



While the formalism for constrained field theories set up by Dirac generalizes to curved backgrounds~\cite{Dirac:1951zz},  the more general formulation in terms of shift and lapse variables was introduced by Arnowitt, Deser and Misner~\cite{Arnowitt:1962hi}. Apart from the gravitational field itself, the ADM decomposition has been used to determine aspects of the Maxwell field on curved backgrounds, which includes the electromagnetic self-energy problem of a point charge~\cite{Arnowitt:1960zza}, the behaviour of the fields near the horizons of stationary black hole spacetimes~\cite{MacDonald:1982zz}, and its quantization~\cite{Sorkin:1979ja}. Through the works of Isenberg and Nester, the ADM decomposition has been subsequently used to better understand the initial value problem of fields theories~\cite{Isenberg:1977ja} and the description of derivative coupled theories ~\cite{Isenberg:1977fs}. While the relevance of boundary terms in the description of the gravitational field has also been considered in ~\cite{Isenberg:1981fa}, the general formulation of constrained theories with boundaries has not been provided in these works. 

Boundaries lead to the inclusion of surface terms which are essential in the formulation of the action in gravity~\cite{Regge:1974zd, Gibbons:1978ac, York:1986} and other field theories~\cite{Gervais:1976ec, Anco:2001gm, Anco:2001gk,Romero:2002vg, G.:2013zca, G.:2015uda}. The generalization to cases where the boundaries are null~\cite{Parattu:2015gga, Booth:2001gx} 
as well as non-orthogonal~\cite{Hawking:1996ww}, have also been considered in the literature. 
There is also much recent activity regarding the asymptotic symmetries of gauge theories and 
gravity~\cite{Balachandran:2013wsa, Campiglia:2014yka, Strominger:2013lka, He:2014cra,
	Strominger:2013jfa, He:2014laa}, which can be formulated in terms of fields localized
on the null boundary of a conformally compactified asymptotically flat 
spacetime~\cite{Mohd:2014oja}. 
While surface terms in these contexts provide an important topic of investigation in its own 
right~\cite{Donnelly:2016auv}, it is not what our paper seeks to address. 

We will be largely concerned in the role 
boundaries play in the classical 
and quantum descriptions of constrained field theories. There has been some consideration of  
these in the literature. The modifications of constraints through the presence of boundaries, 
close to the spirit in which we will carry out our work,  has been investigated in~\cite{SheikhJabbari:1999xd,Zabzine:2000ds}. In~\cite{Balachandran:1993tm,Balachandran:1992qg}, by studying the quantization of the the Chern-Simons theory on a disk, the role of boundaries on the vacuum structure of the theory has been covered in detail. While these works have made some ground in addressing how boundaries affect constraints,  many questions still remain open. As far as we are aware of, there appears to be no general prescription on how boundaries are to be considered in the case of constrained theories, and a formulation on curved backgrounds with horizons is completely lacking. The present work is an attempt to address this issue.

The organization of our paper is as follows. In Sec.~\ref{geom}, we describe the foliation 
which will be implemented to carry out the 3+1 decomposition of the spherically 
symmetric spacetime, as well as the form of the matter action defined on it. In 
Sec.~\ref{Max}, we consider the specific example of Maxwell's electrodynamics
as a constrained theory, for which the concrete manifestation of horizons in the 
description of the dynamics of the fields are pointed out as they arise. Finally, in
Sec.~\ref{Disc} we discuss some unexpected results and possible applications of 
our findings. 

\section{General Algorithm} \label{geom}


We will work on a static, spherically symmetric spacetime endowed with at least one horizon. 
In other words, we assume that there is a timelike Killing vector field $\xi^a\,$ 
with norm given by $\xi^a\xi_a = -\lambda^2\,,$ satisfying 
%
%
\begin{align}
\xi_{[a}\nabla_b \xi_{c ]} & = 0 \,. \label{Frobenius} 
\end{align}
The horizon is defined by $\xi^a$ becoming null, $\lambda=0\,.$ 
It follows that there is a spacelike hypersurface $\Sigma$ which is 
everywhere orthogonal to $\xi^a\,.$ The situation we have in mind is that 
of fields living on the background of a 
static black hole. For an asymptotically flat or anti-de Sitter 
space, $\Sigma$ is the region `outside the horizon', while for a positive 
cosmological constant, we may have a static de Sitter black hole spacetime,
in which case $\Sigma$ is the region `between the horizons'. 

The induced metric on $\Sigma$ is given by
\begin{equation}
h_{ab} = g_{ab} + {\lambda}^{-2} \xi_{a} \xi_{b} \,,
\label{gen.met}
\end{equation}
leading to the following expression for the determinant of spacetime metric
\begin{equation}
\sqrt{-g} = \lambda \, \sqrt{h} \, . \label{gen.det}
\end{equation}

The action functional for $N$ fields $\Phi_A\,,\, A = 1, \cdots, N\,,$ is given by the time 
integral of the Lagrangian $L$, or equivalently the integral of 
the Lagrangian density ${\cal L}$ over the four volume,
\begin{equation}
S[\Phi_A] = \int dt\, L  
= \int dt \int \limits_{\Sigma} \lambda dV_x ~ {\cal L}(\Phi_A(x) , \nabla_a \Phi_A(x)) \, ,
\label{gen.act}
\end{equation}
%
where $dV_x$ is the volume element on $\Sigma\,,$ 
and ${\cal L}(\Phi_A(x) , \nabla_a \Phi_A(x))$ is the Lagrangian density.  The Lagrangian density 
can be written in terms of the `spatial' and `temporal' derivatives of the fields, 
\begin{equation}
{\cal L} \equiv {\cal L}(\Phi_A(x),  \mathcal{D}_a \Phi_A(x),  \dot{\Phi}_A(x) )\,,
\label{gen.den}
\end{equation}
where $\mathcal{D}_a \Phi_A = h_a^b \nabla_b \Phi_A$ are the $\Sigma$-projected derivatives 
of the fields $\Phi_A\,,$ and $\dot{\Phi}_A$ are their time derivatives, defined as their Lie derivatives
with respect to $\xi\,,$
\begin{equation}
\dot{\Phi}_A := \pounds_{\xi} \Phi_A \,.
\label{gen.dot}
\end{equation}
%

The momenta $\Pi^A$ canonically conjugate to the fields $\Phi_A\,$ are defined as
\begin{equation}
\Pi^A  = \frac{\delta L}{\delta \dot{\Phi}_A}  =  -\lambda^{-1}~ \xi_a~ \frac{\partial {\cal L}}{\partial (\nabla_a \Phi_A)} \, ,
\label{gen.mom}
\end{equation}
where the functional derivative in this definition is taken on the hypersurface $\Sigma\,,$ i.e. 
it is an `equal-time' functional derivative, defined as
%
\begin{equation}
\frac{\delta\Phi_A(\vec{x}, t)}{\delta\Phi_B(\vec{y}, t)} = \delta^B_A\, \delta(x, y) = 
\frac{\delta\dot\Phi_A(\vec{x}, t)}{\delta\dot\Phi_B(\vec{y}, t)}\, .
\label{gen.var}
\end{equation}
The $\delta(x,y)$ in Eq.~(\ref{gen.var}) is the three-dimensional covariant delta function defined 
on $\Sigma\,,$
\begin{equation}
\int\limits_\Sigma dV_y \delta(x,y) f(\vec{y}, t) = f(\vec{x}, t)\,.
\label{gen.del}
\end{equation}

Given a Lagrangian $L$ we can construct the canonical Hamiltonian through the Legendre transform
\begin{equation}
H_C = \int \limits_{\Sigma} dV_x ~(\Pi^A \dot{\Phi}_A) -  L \,.
\label{gen.Ham}
\end{equation}
Dynamics in the Hamiltonian formalism is determined using the Poisson bracket, which for two 
functionals $F(\Phi_A(x), \Pi^A(x))$ and $G(\Phi_A(x), \Pi^A(x))$ of the fields and their momenta
is defined as
\begin{equation}
\left[F ,G\right]_P = \int dV_z \left[\frac{\delta F}{\delta \Phi_A(z)} \frac{\delta G}{\delta \Pi^A(z)} 
- \frac{\delta G}{\delta \Phi_A(z)} \frac{\delta F}{\delta \Pi^A(z)} \right]\,. \label{gen.PB}
\end{equation}
The canonical Poisson brackets between the fields and their momenta follows from setting  
$F(x) = \Phi_A(\vec{x}, t)$ and $G(y) = \Pi^B(\vec{y}, t)$
\begin{equation}
\left[\Phi_A(\vec{x}, t), \Pi^B(\vec{y}, t) \right]_P = \delta_{A}^{B} \delta(x,y)\,. 
\label{gen.can}
\end{equation}
The time evolution of any functional of the fields and momenta is determined 
from its Poisson bracket with the Hamiltonian
\begin{equation}
\dot{F}(x) = \left[F(x), H_C \right]_P\,.
 \end{equation}

The Hamiltonian, obtained by a Legendre transform from the Lagrangian,
provides a complete description of the dynamics of the system only if
all velocities of the theory uniquely map into momenta by
Eq.~(\ref{gen.mom}). In the case of constrained theories,  
such a mapping is not possible due to the presence of constraints. In these theories, the Hamiltonian 
must be constructed by determining all the constraints of the theory via the Dirac-Bergmann 
algorithm. The usual Poisson brackets of these theories are modified in the presence of 
constraints, and as we will argue below, the constraints of the theory are 
modified in the presence of horizons.


\section{The Maxwell field}\label{Max}
%
%
For the sake of concreteness, we will consider the specific example of electromagnetism 
as a constrained theory on spherically symmetric spacetimes with horizon(s). The action is
\begin{equation}
S_{EM} =  \int dV_4 \left(-\tfrac{1}{4} F_{a b} F_{c d} g^{a c} g^{b d}\right) \, ,
\end{equation}
%
%
%
%
%
where $dV_4 = \lambda dV_x$ is the four dimensional volume form on the manifold $\Sigma\times\mathbb{R}$, and 
$F_{a b} = 2 \partial_{[a}
A_{b]}$. Defining $e_a = - \lambda^{-1} \xi^c F_{c d}$ and $f_{a b} = F_{c d} h^c_a h^d_b\,,$ we 
can rewrite this action as 
%
\begin{equation}
S_{EM} = - \int dt \int \limits_{\Sigma} dV_x \frac{\lambda}{4} \left[f_{a b} f^{a b} - 2 e_{a} e^{a} \right] \, ,
\label{H.Lag}
\end{equation}
Recalling Eq.~(\ref{gen.dot}), we write 
\begin{align}
\dot A_b \equiv \pounds_{\xi} A_b  &= \xi^{a}\nabla_a A_b + A_a \nabla_a \xi^a \, \notag \\
&=\xi^a F_{a b} + \nabla_a (A_b \xi^b) \,, 
\end{align}
and defining 
	{$\phi = A_a \xi^{a}$},
	we have for the projection $a_b\,,$
%
%
\begin{equation}
\dot a_b = -\lambda e_b + \mathcal{D}_b \phi \,.
\label{H.elf}
\end{equation}
%
Since the velocity term $\pounds_{\xi} \phi$ does not appear in the electromagnetic Lagrangian Eq.~(\ref{H.Lag}), it implies that the momentum conjugate to  $\phi$ vanishes, 
\begin{equation}
\frac{\partial L_{E M}}{\partial \dot{\phi}} = \pi^{\phi} = 0 \,,
\label{H.con}
\end{equation}
thus producing the only constraint following from the Lagrangian. The momenta 
corresponding to the $a_b$ are given by
%
\begin{equation}
\pi^b = \frac{\partial L_{E M}}{\partial \dot{a}_b} = - e^b \, .
\label{H.mom}
\end{equation} 

The canonical Hamiltonian is then 
%
%
		\begin{align} 
H_C &= \int \limits_{\Sigma} dV_x \, \left(\pi^b \dot{a}_b\right) -  L \notag \\
&= \int \limits_{\Sigma} dV_x \, \left[\lambda\left(\frac{1}{2} \pi^b \pi_b + \frac{1}{4} f_{a b} f^{a b}\right) + \pi^b \mathcal{D}_b \phi \right]\,.
\end{align}
%
%
The constraint of Eq.~(\ref{H.con}) is now added to this and a new Hamiltonian is defined,
%
%
		\begin{equation}
H_0  = \int \limits_{\Sigma} dV_x \, \left[\lambda \left(\frac{1}{2} \pi^b \pi_b + \frac{1}{4} f_{a b} f^{a b}\right) + \pi^b \mathcal{D}_b \phi + v_{\phi} \pi^{\phi} \right] \, ,
\label{H.pri}
\end{equation}
where $v_{\phi}$ is an undetermined multiplier. The canonical Poisson brackets of
Eq.~(\ref{gen.can}) are in this case
\begin{align}
\left[\phi(x), \pi^{\phi}(y) \right]_P & = \delta(x,y) \notag\\
\left[ a_a(x) , \pi^b(y) \right]_P & = \delta^{b}_{a}\delta(x,y) \,.
\label{H.can}
\end{align}
%
%
\subsection{The Dirac-Bergmann algorithm} \label{MDBA}
We will now apply the Dirac-Bergman algorithm to determine all the constraints of the theory
and construct the unconstrained Hamiltonian. 
%
For that, we need to check that the constraint is obeyed at all times, or in other words, 
$\dot \pi^{\phi} \approx 0 \,.$ The Poisson bracket between $\pi^\phi$ and the Hamiltonian 
is calculated with the help of a smearing function $\epsilon$ as follows,
\begin{align}
\int \limits_{\Sigma} dV_y \epsilon(y) \dot\pi^\phi(y) &= \int \limits_{\Sigma} dV_y 
\epsilon (y) \left[\pi^{\phi}(y) , H_0 \right]_P  \,\notag \\
&= \int \limits_{\Sigma} dV_y 
\epsilon (y) \left[\pi^{\phi}(y) , \int \limits_{\Sigma} dV_x \pi^b(x) 
\mathcal{D}^x_b \phi(x) \right]_P \notag \\
&=  - \oint \limits_{\partial \Sigma} da_y \, \epsilon(y)  n^y_b \pi^b(y) + 
\int  \limits_{\Sigma} dV_y \, \epsilon(y) \left( \mathcal{D}^y_b \pi^b(y)\right)\,.
\label{U.PB1}
\end{align}
Here we have used the canonical Poisson brackets given in Eq.~(\ref{H.can}) 
and an integration by parts. The smearing function $\epsilon$ is assumed
to be well behaved, but we make no further assumption regarding its properties. 
In particular we do not assume that $\epsilon$ vanishes on the 
horizon (or horizons, if $\Sigma$ is the region between the horizons in a de Sitter 
black hole spacetime). 

%
%
The surface integral is finite, since using Schwarz inequality we get
\begin{equation}
\left| n_b \pi^b \right|  \leq \, \sqrt{\left|n_b n^b\right| \, 
	\left|\pi_b \pi^b\right|}  \,.
\label{U.SI}
\end{equation}
In this, $n_bn^b = 1$ by definition since $n$ is the `unit normal' to the horizon,
and $\pi_b\pi^b = e_b e^b$ appears in the energy momentum tensor (more precisely
in invariant scalars such as $T^{ab}T_{ab}$), and therefore may not
diverge at the horizon.
So the integral over $\partial\Sigma$ is finite and in general different from zero.
Thus the boundary integral is finite at the horizon and we have a non-vanishing 
contribution from $\partial\Sigma$\,, which was one of the things we wanted to show.
%

We note that since the smearing function $\epsilon$ is present in the integrand,
specific assumptions about the class of allowed smearing functions may be required to 
produce physically sensible results. For the Maxwell case, the assumption 
that $\epsilon$ is regular at the horizon (with no dependence on $\lambda$) 
is sufficient.

The right hand side of Eq.~(\ref{U.PB1}), comprising of a bulk and a surface term,
must vanish weakly, giving a constraint. Since we are working on a spherically symmetric 
background, we can use a radial delta function to convert the surface integral to
a volume integral,
\begin{equation}
\oint \limits_{\partial \Sigma} da_y  K(y) = \int  \limits_{\Sigma} dV_y \lambda(y) K(y) 
\delta(r(y) - r_H) \,,
\label{U.area}
\end{equation}
where $K$ is any well-behaved function, $r_H$ is the radius of the sphere corresponding 
to $\partial\Sigma\,,$ $\delta$ 
is the usual Dirac delta, defined by $\int\, dr\, \delta(r - R) f(r) = f(R)$ for 
any well-behaved function $f(r)\,,$ and we have assumed that $h^{rr} = \lambda$ for the 
spherically symmetric metrics that we consider. (This is solely for notational convenience,
and $h^{rr}$ should replace $\lambda$ in this formula if there is any confusion.) If we have 
a de Sitter black hole spacetime, we will need to consider a sum over two spheres,
corresponding to inner and outer horizons. 
Thus we can rewrite the last equality in Eq.~(\ref{U.PB1}) as
%
%
%
%
%
\begin{equation}
\int \limits_{\Sigma} dV_y \epsilon(y) \dot\pi^\phi(y) 
= \int  \limits_{\Sigma} dV_y \, \epsilon(y) \left[ - \lambda(y)  n^y_b \pi^b(y) 
\delta (r(y) - r_H) + \mathcal{D}^y_b \pi^b(y) \right]\,,
\label{U.PB11}
\end{equation} 
which produces the constraint
%
%
%
\begin{equation}
\Omega_2 =  - \lambda n_b  \pi^b\delta (r - r_H) + \mathcal{D}_b \pi^b \approx 0 \, .
\label{U.con2}
\end{equation}

We now need to check if there are any further constraints resulting from $\dot \Omega_2 \approx 0$. We first include the new constraint with a multiplier into the existing Hamiltonian given in Eq.~(\ref{H.pri}), which gives us 
%
%
%
\begin{equation}
H_T = H_0 + \int \limits_{\Sigma} dV_x v_1 \left[\mathcal{D}_b  \pi^b - \lambda n_b \pi^b\delta (r - r_H) \right] \,,
\label{U.sham}
\end{equation}
and consider the time evolution to be governed by this Hamiltonian,
		\begin{align}
\int \limits_{\Sigma} dV_y\, \epsilon(y) \dot \Omega_2(y) &= \int \limits_{\Sigma} dV_y\,
\epsilon(y)  \left[ \Omega_2(y) , H_T \right]_P \notag \\
&= - \int \limits_{\Sigma} dV_y  \mathcal{D}^y_b\left( \epsilon (y)\right) \int \limits_{\Sigma} dV_x \left[ \pi^b(y) , \mathcal{D}^x_a a_c(x) \right]_P f^{a c}(x) \notag \\
&= \int \limits_{\Sigma} dV_y  \mathcal{D}^y_b \left( \epsilon(y)\right) \int \limits_{\Sigma} dV_x \mathcal{D}^x_a \left(\delta(x,y)\right) f^{a b}(x)  \notag\\
&= \int \limits_{\Sigma} dV_x \mathcal{D}^x_a \left(\int \limits_{\Sigma} dV_y  \mathcal{D}^y_b \left( \epsilon(y)\right)\delta(x,y)\right) f^{a b}(x)  \notag\\ 
& = \int \limits_{\Sigma} dV_y   \mathcal{D}^y_a \mathcal{D}^y_b \epsilon(y) f^{a b}(y) \notag\\ 
& = 0\,.
\label{U.PB2}
\end{align}
The last equality follows from the antisymmetry of $f^{ac}$ in its indices, and we have used the fact that 
$\mathcal{D}$ is torsion-free. Since these constraints commute with one another, they are also first class. 
Thus the full Hamiltonian is the $H_T$ defined earlier,
%
%
%
\begin{equation}
H_T = \int\limits_{\Sigma} dV_x \left[ \lambda \left( \frac{1}{4} f_{a b} f^{a b} + \frac{1}{2} \pi_a \pi^a\right) + v_1 \left( \mathcal{D}_b \pi^b -  \lambda n_b \pi^b \delta (r - r_H) \right) + \pi^b \mathcal{D}_b \phi + v_{\phi} \pi^{\phi} \right]
\label{U.Htot1}
\end{equation}
The multipliers $v_1$ and $v_\phi$ may be determined by examining the equations of motion. 
The evolution of $\phi$ is given by
\begin{align}
\int \limits_{\Sigma} dV_y \epsilon(y) \dot \phi(y) &= \int \limits_{\Sigma} dV_y 
\epsilon(y)\left[  \phi(y) , H_T \right]_P \notag \\
&=  \int \limits_{\Sigma} dV_y \epsilon(y) \int \limits_{\Sigma} dV_x v_{\phi}(x) 
\left[ \phi(y) , \pi^{\phi}(x) \right]_P \notag \\
&=  \int \limits_{\Sigma} dV_y \epsilon (y) v_{\phi}(y) \, ,
\end{align}
which tells us that we can set $\dot \phi = v_{\phi}$. The evolution of $a_b$ can also 
be determined in the same manner,
%
 %
\begin{align}
\int \limits_{\Sigma} & dV_y\, \epsilon(y) \dot{a}_b(y) = \int \limits_{\Sigma} dV_y\, \left[ \epsilon(y) a_b(y) , H_T \right]_P \notag \\
&= \int \limits_{\Sigma} dV_y\, \epsilon(y)  \int \limits_{\Sigma} dV_x \left[a_b(y) , \pi^c(x) \right]_P \left(\lambda(x) \pi_c(x) + \mathcal{D}^x_c\phi(x) - \mathcal{D}^x_c v_1(x)\right)  \notag \\
&=  \int \limits_{\Sigma} dV_y \epsilon(y) \left[ \lambda(y) \pi_b(y) + \mathcal{D}^y_b \phi(y) - \mathcal{D}^y_b v_1(y) \right] \, .
\label{U.vpt}
\end{align}
Comparing this with Eq.~(\ref{H.elf}), we deduce that we can set 
$\mathcal{D}_b v_1 =  0$. There may be many ways in which this 
condition could be satisfied,
but for simplicity, we will simply assume that $v_1 =0$. 
Then Eq.~(\ref{U.vpt}) produces
%
%
\begin{equation}
\dot a_b = \lambda\pi_b + \mathcal{D}_b \phi \, ,
\end{equation}
and we thus find that the total Hamiltonian takes the form
%
%
	\begin{equation}
H_T = \int\limits_{\Sigma} dV_x \left[\lambda \left(\frac{1}{4} f_{a b} f^{a b} + \frac{1}{2} \pi_a \pi^a\right) + \pi^b \mathcal{D}_b \phi + \dot{\phi} \pi^{\phi} \right] \,.
\end{equation}
\subsection{Gauge transformations and Gauge fixing}

We have found two constraints, both first class, which depend on the momenta in the theory. These on account of being first class will generate gauge transformations, i.e. they transform the fields while not transforming the Hamiltonian (or the Lagrangian).


For the constraint $\Omega_1 = \pi^{\phi}$ , the only non-vanishing Poisson Bracket with 
the fields is
\begin{align}
	\delta_1 \phi(y) &= \left[ \phi(y) , \int\limits_{\Sigma} dV_x \epsilon_1 (x) \pi^{\phi}(x) \right]_P \notag \\
	&= \epsilon_1 (y) \,.
	\label{GT1}
\end{align}
%
%
For the other first class constraint $\Omega_2$ of the theory, we have the following
non-vanishing Poisson bracket with the fields
%
%
\begin{align}
	\delta_2 a_b(y) &= \left[ a_b(y) , \int\limits_{\Sigma} dV_x \epsilon_2 (x) \left(\mathcal{D}^x_c \pi^c(x) -  \lambda(x) n^x_c \pi^c(x) \delta(r(x) - r_H) \right) \right]_P \notag \\
	&= \left[ a_b(y) ,  \int\limits_{\Sigma} dV_x \pi^c(x) \mathcal{D}^x_c \epsilon_2(x) \right]_P \notag \\
	&= \mathcal{D}^y_b \epsilon_2(y)\, .
	\label{GT2}
\end{align}
These transformations can be identified with the usual gauge transformations $A_\mu \to A_\mu + \partial_\mu \psi$ if we write
$\epsilon_2(y) = \psi$ and 
$\epsilon_1(y) = \pounds_{\xi} \psi$. This can be seen by 
simply projecting the gauge transformation one finds from the Lagrangian
%
%
\begin{align}
	A_a(x) + \nabla^x_a \psi(x) & = \delta^b_a \left( A_b(x) + \nabla^x_b \psi(x) \right) \notag\\
	&=a_a + \mathcal{D}^x_a \psi(x) -\lambda^{-2}(x) \xi_{a} \left(\phi(x) + \pounds_{\xi} \psi(x) \right) \, .
\label{gf.GTL}
\end{align}
We note that the gauge transformations for this background are the same as in the 
absence of horizons. The boundary terms which arise in the constraints are such that the 
gauge transformations remain unaltered.

To proceed further, we take the approach of converting the gauge constraints into 
second class ones by fixing the gauge. Let us choose a `radiation-like' gauge, in 
which the gauge-fixing condition or constraint is analogous to the usual radiation 
gauge, with an additional boundary contribution motivated by the 
surface term in Eq.~(\ref{U.con2}). This choice considers how the horizon could 
affect the dynamics of the theory, which was our original motivation for this work.
Thus we now have a total of four constraints given by
%
%
\begin{align}
	\Omega_1 & = \pi^{\phi} \notag \\
	\Omega_2 & =  \mathcal{D}_a \pi^a
	- \lambda n_a \pi^a \delta (r - r_H) \notag\\
	\Omega_3 & = \phi \notag \\
	\Omega_4 & = \mathcal{D}^b a_b 
	- \lambda n^b a_b \delta (r - r_H)\,.
	\label{gf.con}
\end{align}
We note that $a_b$ is not a physical observable since it changes under a gauge transformation. 
In particular, the near horizon behaviour of the coefficient of the $\delta$-function in the 
last term of the gauge-fixing constraint\, $\Omega_4$ cannot be fixed from any physical 
consideration. The Poisson brackets of these constraints are easily calculated,
%
%
\begin{align}
	\left[\Omega_1 (x), \Omega_3 (y) \right]_P & = - \delta(x,y) \, , \notag \\
	\left[\Omega_2 (x), \Omega_4 (y) \right]_P & =  
	 \mathcal{D}_{a} \mathcal{D}^{a}\delta(x,y)  \, , 
	\label{gf.pb}
\end{align} 
with all other Poisson brackets vanishing. The first Poisson bracket in 
Eq.~(\ref{gf.pb}) follows directly from the canonical relations. The 
second Poisson bracket gives 
%
%
\begin{align}
	& \left[ \int dV_x \eta(x) \Omega_2 (x) , 
	\int dV_y \epsilon(y) \Omega_4(y) \right]_P \notag\\
	 & \qquad \qquad  = \left[ \int dV_x  \left(\mathcal{D}^x_a 
	 \eta(x) \right)  \pi^a(x)\,, \int dV_y \left(\mathcal{D}_y^b\epsilon(y) \right) a_b(y) \right]_P \notag \\
	& \qquad \qquad  = - \int dV_y  \left(\mathcal{D}^y_a
	\eta(y)\right) \left(\mathcal{D}_y^a\epsilon(y) \right) \notag \\
	& \qquad \qquad  = -\oint da_y \epsilon(y) n_y^a  \left(\mathcal{D}^y_a\eta(y)\right) + \int dV_y \epsilon(y) \mathcal{D}_y^a \mathcal{D}^y_a\eta(y)\,,
	\label{gf.pb2}
\end{align}
where we have used integration by parts in deriving the equalities.
%
The surface integral in the last equality of Eq.~(\ref{gf.pb2}) vanishes, 
which can be seen by using Schwarz's inequality
%
%
\begin{align}
	\left| n^a D_a \left(\eta\right) \right|^2 & \leq \left|n^a n_a \right| \left|h^{ab} \left(D_a\eta\right) \left(D_b\eta\right)\right| \notag\\
	& = h^{a b} \left(D_a\eta\right) \left(D_b\eta\right) \,.
	\label{gf.srf}
\end{align}
The smearing function and its derivatives are regular on the horizon, while 
$h^{rr} \sim \lambda^2$ on spherically symmetric backgrounds. Hence the surface integral 
in the last equality of Eq.~(\ref{gf.pb2}) vanishes, and only the 
volume term contributes. 
%
%
Thus the Poisson brackets between the constraints are those given in Eq.~(\ref{gf.pb}).
%
%
The matrix of the Poisson brackets between these constraints have 
a non-vanishing determinant and is invertible. This matrix, 
$C_{\alpha\beta}\left(x,y \right) = \left[\Omega_\alpha (x), 
\Omega_\beta(y)\right]_P\,,$ is given by
%
%
%
%
%
%
%
%
%
\begin{equation}
	C (x,y) = 
	\begin{pmatrix} 
		0 & 0 & -\delta(x,y)\ & 0 \\
		0 & 0 & 0 & \mathcal{D}_a \mathcal{D}^a\delta(x,y) \\
		\delta(x,y) & 0 & 0 & 0 \\
		0 & - \mathcal{D}_a \mathcal{D}^a\delta(x,y) & 0 & 0
	\end{pmatrix} \, .
	\label{gf.dirm}
\end{equation}
The Dirac bracket for any two dynamical entities $F$ and $G$ (which may be 
functions on the phase space, or functionals, the duals of functions) is defined as 
%
\begin{equation}
	\left[F\,, \,G \right]_{D} = \left[F\,, \,G \right]_{P} - \int dV_w \int dV_z 
	\left[F\,, \,\Omega_\alpha(w)\right]_P C^{-1}_{\alpha\beta}\left(w, z\right) \left[\Omega_\beta(z)\,, \,G\right]_P\,.
	\label{gf.dir}
\end{equation}  
%
%
%
Thus we need to find the inverse of the operator
$\mathcal{D}_a
		\mathcal{D}^a$.
Let us formally write the inverse as $\tilde{G}(x, y)\,,$ i.e. 
%
%
\begin{equation}
	\mathcal{D}_a\mathcal{D}^a \tilde{G}\left(x, y\right) = \delta\left(x,y\right) \,.
	\label{U.gfs}
\end{equation}
This $\tilde{G}(x, y)\,$ is the Green's function 
for the {\em spatial} Laplacian operator $\mathcal{D}_a\mathcal{D}^a $\,, but not of the wave operator, 
which is the {\em spacetime} Laplacian. With the help of this, the inverse matrix 
$C^{-1}_{\alpha\beta}(x,y)$ can be written as

%
%
%
%
%
%
\begin{equation}
	C^{-1}(x,y) = 
	\begin{pmatrix}
		0 & 0 & \delta(x,y) & 0\\
		0 & 0 & 0 & -\tilde{G}\left(x,y\right) \\
		-\delta(x,y) & 0 & 0 & 0 \\
		0 & \tilde{G}\left(x,y \right) & 0 & 0
	\end{pmatrix} \, .
	\label{gf.cin}
\end{equation}
We can now substitute Eq.~(\ref{gf.cin}) in Eq.~(\ref{gf.dir}) to find the following Dirac brackets for the fields,
%
%
%
\begin{align}
	\left[a_a(x),\pi^b(y)\right]_{D} & = \delta(x,y)\delta_a^b - \mathcal{D}_a^x \mathcal{D}^b_x \tilde{G}\left(x,y \right)\,,
\label{gf.cdir}
\end{align}
all other Dirac brackets being zero.

We could choose to fix the gauge so that the resulting Dirac brackets would involve
the (static) Green's function for the spacetime Laplacian. The corresponding gauge-fixing
constraints are
%
%
\begin{align}
	\Omega_3 & = \phi \, ,\notag \\
	\Omega_4 & = \mathcal{D}^b\left(\lambda a_b\right) \, .
	\label{gf.con2}
\end{align}  
In this gauge the Dirac brackets are given by
\begin{align}
	\left[a_a(x),\pi^b(y)\right]_{D} & = \delta(x,y)\,\delta_a^b -  
	\mathcal{D}_a^x\left(\lambda(x)\mathcal{D}^b_x G\left(x,y \right)\right)\,,
\label{gf.cdir2}
\end{align} 
all other Dirac brackets being zero. Here $G(x, y)$ is the time-independent Green's function 
for the spacetime Laplacian,
\begin{equation}
	\mathcal{D}_a^x\left(\lambda(x)\mathcal{D}^a_x G\left(x,y \right)\right) = \delta(x,y)\,.
\end{equation}

Although the choice of gauge-fixing functions determine the form of Dirac brackets, we know 
that physical observables and measurable quantities must be independent of that choice. 
However, the choice of Green's function is determined by the boundary conditions
we wish to impose on the fields at the horizon. We have mentioned earlier that 
the horizon is not a boundary of the spacetime, in particular it is not necessary to 
impose boundary conditions which force the respective fields to vanish on the
horizon, or even remain finite on the horizon if we are considering gauge fields. 
We expect that the utility of our choice of gauge-fixing in Eq.~(\ref{gf.con}) could lie 
in its allowance for more general boundary conditions for gauge fields at the horizon, 
and in its ability to describe the dynamics of fields at the horizon. We will leave a
detailed investigation for later work, but discuss one interesting result, as well
as some open questions, in the next section.

We note in passing that explicit forms of both kinds of Green's functions considered in this section are known for the Schwarzschild 
black hole as well as for the static de Sitter backgrounds~\cite{Cop:1928, J.Math.Phys.12.1845, Hanni:1973fn, Linet:1976sq, Bunch:1977sq, Chernikov:1968zm, Tagirov:1972vv, Dowker:1977, Fernandes:2016sue}. 

\section{Discussion}\label{Disc}
In this work, we have argued that horizons affect the constraints of gauge theories, and 
will in turn affect their dynamics in ways that spatial boundaries cannot. In the previous
section, we demonstrated that the Gauss law constraint of the Maxwell field receives a surface 
contribution on spherically symmetric backgrounds with horizons. This however will not 
typically happen for backgrounds with spatial boundaries. While the behaviour of gauge 
fields at spatial boundaries cannot be determined by physical considerations alone, they must
nevertheless be continuous across them.  When surface terms occur, they are present on either 
side of the spatial boundary and cancel out. When the boundary corresponds to the physical end 
of the manifold, regularity of the fields require that they vanish there. The surface terms of 
the previous section exist because the horizon hides the `other' side of the horizon from
observations. The only requirement are that physical fields, more precisely gauge-invariant 
(and local Lorentz-invariant) scalars constructed out of the fields, must be finite at the 
horizon. Horizons thus lead to a richer set of possibilities for field theories, 
and in particular for gauge theories.
%
%

%
We have also considered the gauge fixing of the Maxwell field as a specific example to test 
our claim. We can get some insight into the role played by the horizon in the choice of
boundary conditions by comparing the two gauge fixing constraints of Eq.~(\ref{gf.con}) and 
Eq.~(\ref{gf.con2}). These gauge choices led to Dirac brackets which involve the 
Green's functions for the spatial Laplacian and the spacetime Laplacian, respectively. The 
difference in the actions of the two Laplacian operators on a time-independent 
scalar field $F$ is seen in the following identity,
\begin{equation}  
\nabla_a\nabla^a F - \mathcal{D}_a \mathcal{D}^a F =  \left(\lambda^{-1} \mathcal{D}^a \lambda\right) D_{a} F \, .
\label{App.Lap}
\end{equation}
When $\lambda=\lambda(r)$, as is the case for spherically symmetric backgrounds, the limit of Eq.~(\ref{App.Lap}) as $r \to r_H$ is
%
\begin{equation}
\left[\nabla_a\nabla^a F - \mathcal{D}_a \mathcal{D}^a F\right]_{r=r_H} = \kappa_H \left(\partial_r F\right)_{r=r_H} \, .
\label{App.diff}
\end{equation}
Thus the action of the spacetime and spatial Laplacians differ at the horizon by a term which depends on the surface 
gravity $\kappa_H$ of the background. Eq.~(\ref{App.diff}) also indicates how boundary conditions affect the behaviour 
of the operators on the left hand side. For instance, 
the operators disagree on the horizon when either Dirichlet or Robin boundary conditions are assumed. 
This suggests that the eigenvalues, and thus the description of the horizon states may be different for the two gauge choices,
however that is an investigation we leave for another occasion.

The modified Gauss law constraint in Eq.~(\ref{U.con2}) has further implications on the quantization of this theory, including horizon states and the charge. By integrating this constraint over a bounded volume whose radius is greater than the event horizon of the black hole (and less than the cosmological horizon, should it exist), we can derive the expression for the charge contained within it. In considering the Reissner-Nordstr\"om solution for simplicity, the electric field given in Eq.~(\ref{H.elf}) reduces to $\lambda \pi_b = - \mathcal{D}_b \phi$. Integrating up to a spatial boundary of radius $r_B$, where $r_B > r_H$, we have 
\begin{align}
Q &=  \int \limits_{\Sigma_B} \Omega_2 \notag\\
&= - \oint \limits_{\partial_{\Sigma}} \left[r^2\partial_r \phi(r) \right]_{r = r_B} +  \oint \limits_{\partial_{\Sigma}} \left[r^2\partial_r \phi(r) \right]_{r = r_H} -  \oint \limits_{\partial_{\Sigma}} \left[r^2\partial_r \phi(r) \right]_{r = r_H} \notag\\
&= - \oint \limits_{\partial_{\Sigma}} \left[r^2\partial_r \phi(r) \right]_{r = r_B} \, .
\label{App.charge}
\end{align}
In Eq.~(\ref{App.charge}), `$\Sigma_B$' indicates that the volume integral on the hypersurface is evaluated from the horizon up to a sphere of constant radius $r_B$. The usual spherically symmetric solution, $\phi = \frac{Q}{4 \pi r}$, satisfies this equation. A crucial difference occurs if the above integral is performed \emph{exactly} at the black hole horizon, in which case we have
\begin{align}
Q_H &= \int \limits_{\Sigma_H} \left[\mathcal{D}_a \pi^{a} - \lambda n_a \pi^a \delta(r - r_H) \right] \notag\\
&=0\,.
\label{App.nocharge}
\end{align}
Note that Eq.~(\ref{App.nocharge}) is the charge at the horizon(s) and holds regardless of the solution for the charge outside the horizon. Equations (\ref{App.charge}) and (\ref{App.nocharge}) suggests that the horizon may be viewed as a dipole layer, with the charge on one side of the horizon being screened from observation. For an observer outside the horizon, the black hole is a charged body which follows from the bulk contribution to the constraint. When the observer is at the horizon, the cancellation of ``positive" and ``negative" charges leads to the result in Eq.~(\ref{App.nocharge}). 

%
The constraint in Eq.~(\ref{U.con2}) will also necessarily affect the quantization of the Maxwell field on such backgrounds. Quantization of gauge fields is most effectively carried out within the BRST formalism, where the first class constraints of the theory and the inclusion of additional ghost fields leads to the construction of the BRST charge operator. Within the Hamiltonian BRST formalism, this operator leads to the derivation of the gauge fixing and ghost actions at tree level, both of which are BRST invariant \cite{Henneaux:1985kr,Henneaux:1992ig}.  On backgrounds with boundaries, the requirement of BRST invariance imposes certain restrictions on the allowed boundary values of the fields involved. Such considerations have been made on curved backgrounds with spatial boundaries 
in~\cite{Moss:1989wu,Moss:1990yq,Moss:1996ip,Moss:2013vh}, and more recently in~\cite{Huang:2014pfa,Donnelly:2014fua,Donnelly:2015hxa} in relation to edge state entanglement
entropy calculations. These investigations however did not consider the modification of the Gauss law constraint. That the constraint derived in Eq.~(\ref{U.con2}) does contain a non-vanishing contribution at the horizon should lead to the derivation of ``horizon terms" for the action, which are guaranteed to be BRST invariant. Further, it is clear that when Eq.~(\ref{U.con2}) is imposed as an operator on physical states, it will relate the bulk and horizon states in a non-trivial way. To our knowledge, such considerations have not been made in the literature, and could be particularly relevant in describing the behaviour of horizon states of gauge fields. 

%
To summarize, we have presented a covariant formalism for describing constrained field theories in 
the presence of an event horizon of spherically symmetric black hole spacetimes, which are 
either asymptotically flat or have an outer cosmological horizon. In the process we also 
determined that the presence of horizons lead to non-trivial surface contributions to the 
constraints in general, and demonstrated this explicitly in the Maxwell case. We have argued that surface contributions to the constraints will modify both the quantization of these theories and the description of states at the horizon. We leave further investigation of these topics to future work.

\section*{Acknowledgments}

KF thanks Steve Carlip, Daniel Kabat, Aruna Kesavan, George Paily and Robert Wald for stimulating discussions.


\end{document}